\documentclass[twocolumn,showpacs,floats,floatfix,superscriptaddress,aps,pra]{revtex4}
\usepackage{eurosym}
\usepackage{amsfonts}
\usepackage{amssymb}
\usepackage{amsmath}
\usepackage{graphicx}
\usepackage{bm}
\usepackage{color}
\usepackage{epsfig}
\usepackage{calc}
\usepackage{ifthen}
\usepackage{blindtext}
\usepackage{xcolor}
\usepackage{subfigure}
\usepackage{physics}

\usepackage[hidelinks, colorlinks=true,linkcolor=blue,citecolor=blue,filecolor=blue,urlcolor=blue]{hyperref}

\begin{document}
\title{Four-wave plate composite polarization controller}
\date{\today }

\begin{abstract}
We theoretically propose and experimentally demonstrate a novel composite polarization controller. With our design, which comprises two half-wave plates and two quarter-wave plates the retardance and rotation can be changed continuously by simply rotating the half-wave plates. The idea is universal since any commercial half and quarter-wave plates may be used to achieve any desired polarization change. 
\end{abstract}


\author{Hristina Hristova}
\affiliation{Institute of Solid State Physics, Bulgarian Academy of Sciences, 72 Tsarigradsko chauss\'{e}e, 1784 Sofia, Bulgaria}
\author{Hristo Iliev}
\email{h_iliev@phys.uni-sofia.bg}
\affiliation{Faculty of Physics, Sofia University, James Bourchier 5 Blvd, 1164 Sofia, Bulgaria}
\author{Ivayla Bozhinova}
\affiliation{Faculty of Physics, Sofia University, James Bourchier 5 Blvd, 1164 Sofia, Bulgaria}
\author{Andon Rangelov}
\affiliation{Center for Quantum Technologies, Faculty of Physics, Sofia University, James Bourchier 5 Blvd, 1164 Sofia, Bulgaria}
\author{Asen Pashov}
\affiliation{Faculty of Physics, Sofia University, James Bourchier 5 Blvd, 1164 Sofia, Bulgaria}

\maketitle

\section{Introduction}
Since polarization is one of the fundamental properties of light \cite{Goldstein,Damask} the ability to monitor and manipulate it is required for many practical applications. Optical polarization retarders and rotators are the two primary optical devices used to control the polarization state. 

\color{black}
In general, a retarder is an optical device that introduces a phase difference between the projections of the incoming wave field vector $\vec{E}$ onto the two orthogonal directions, called fast and slow axes \cite{Goldstein,Damask,Ardavan,Peters,Lyu}. As a result, depending on the phase shift and the amplitudes of the components (i.e. the mutual orientation of the wave plate and the input $\vec{E}$ vector), different output polarization states can be achieved. In particular, if the phase shift is $\pi$ (i.e. $\lambda/2$ plate), then the retarder transforms the linear input polarization again into linear but rotated at some angle.  If the phase difference is $\pi/2$ (i.e. $\lambda/4$ plates) then the retarder transforms the linear input polarization into elliptical, with ellipticity parameters that depend on the orientation of the input polarization and the wave plate.

A polarization rotator, on the other hand, is an optical device that modifies the polarization state of an incoming light wave by rotating its polarization plane by a defined angle, regardless of its initial polarization orientation.  
\color{black}

From general considerations, a combination of two quarter-wave plates with a half-wave plate between them can map arbitrary input states to an arbitrary output state \cite{Damask}, but this configuration simply cascades the operation of the three wave plates and requires adjustment of their orientation for each desirable output. Recently, Messaadi et al. \cite{Messaadi}  proposed a wave plate with adjustable retardance. Their simple approach is built on two half-wave plates placed between two quarter-wave plates. One of the half-wave plates in this optical system may be rotated to tune the retardation. 

\color{black}
In this paper, we introduce a novel polarization controller, which combines the same elements as in Ref.~\cite{Messaadi} but rearranged to realize a device that can be used simultaneously as an arbitrary retarder and an arbitrary rotator. \color{black} Since any reversible polarization transition can be seen as a composition of a retarder and rotator \cite{Hurvitz}, by using our device it is possible to convert polarization from any initial state to any desired output state. Contrary to the universal device from \cite{Damask}, only two elements need to be adjusted. 
In Section~\ref{theory} we present the theoretical background behind our device and discuss the conditions under which it can be considered as universal. In section~\ref{experiment} we describe the experimental realization and estimate the accuracy of its performance.  In section~\ref{simulation} we present simulations which visualize the range of possible output polarization states for various inputs.

\begin{figure*}[thb]
\centerline{\includegraphics[width=0.9\textwidth]{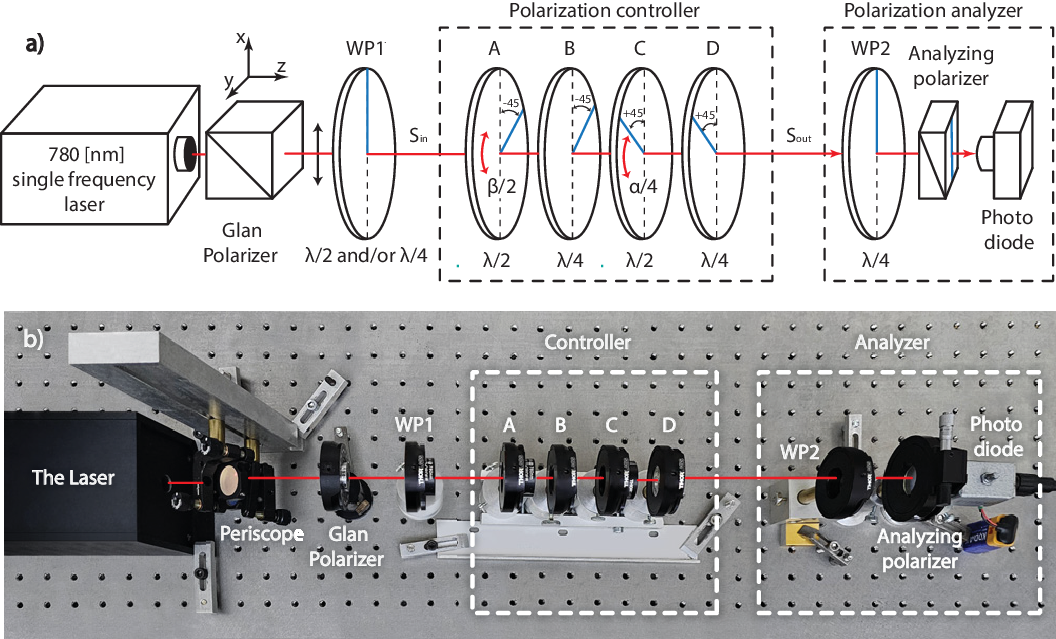}}
\caption{{(Color online) Experimental setup a) a schematic drawing where, WP1 and WP2 - wave plates, $S_{\mathrm{in}}$ - input Stokes vector, $S_{\mathrm{out}}$ - output Stokes vector and b) photo image of the experimental implementation}}
\label{Setup}
\end{figure*}

\section{Theory}
\label{theory}

The Jones matrix describing a rotator with a rotation angle $\theta$ is given by \cite{Goldstein,Damask}:
\begin{equation}
\mathbf{R}(\theta) = \left[
\begin{array}{cc}
\cos \theta & \sin \theta \\
-\sin \theta & \cos \theta
\end{array}
\right],
\end{equation}
while the Jones matrix representing a retarder is \cite{Goldstein,Damask}:
\begin{equation}
\mathbf{J}(\varphi) = \left[
\begin{array}{cc}
e^{i\varphi/2} & 0 \\
0 & e^{-i\varphi/2}
\end{array}
\right],
\end{equation}

\noindent where $\varphi$ is the phase shift between the orthogonal components. 
Now, consider a single wave plate with a phase shift $\varphi$, whose slow and fast axes are rotated by an angle $\theta$ relative to the horizontal and vertical axes. The Jones matrix of the rotated wave plate $\mathbf{J}_{\theta}(\varphi)$ is given by
\begin{equation}
\mathbf{J}_{\theta}(\varphi) = \mathbf{R}(-\theta) \mathbf{J}(\varphi) \mathbf{R}(\theta).  \label{retarder}
\end{equation}

To examine a basic series of two half-wave plates with a relative rotation angle $\beta/2$ between their fast polarization axes, we multiply the Jones matrices for the two half-wave plates specified in Eq. (\ref{retarder}), where the rotation of each wave plate from the horizontal-vertical axes are $\theta$ and $\theta + \beta/2$. Thus, the total propagator is

\begin{equation}
\mathbf{J}_{\theta}(\pi) \mathbf{J}_{\theta + \beta/2}(\pi) = -\left[
\begin{array}{cc}
\cos \beta & \sin \beta \\
-\sin \beta & \cos \beta
\end{array}
\right],  \label{rotator}
\end{equation}

\noindent which is the Jones matrix for rotator  (up to a $\pi$ phase) \cite{Zhan}. We note that angle $\theta$ is an irrelevant free parameter, which means that a rotator rotated additionally at angle $\theta$ is the same rotator. Furthermore, using the well-known fact that a rotator placed between two crossed quarter-wave plates acts as an arbitrary retarder with retardation equals to two times the rotator's rotation angle \cite{Ye},   we construct a tunable retarder in the same manner as in \cite{Messaadi}:

\begin{equation}
\mathbf{J}_{0}(\alpha) = \mathbf{J}_{-\frac{\pi}{4}}(\pi/2) \mathbf{J}_{\theta_1}(\pi) \mathbf{J}_{\theta_1 + \frac{\alpha}{4}}(\pi) \mathbf{J}_{\frac{\pi}{4}}(\pi/2).
\end{equation}

Now, we combine this tunable retarder with the rotator from Eq. (\ref{rotator}), we get

\begin{gather}
\mathbf{J} = \mathbf{J}_{0}(\alpha) \mathbf{R}(\beta) =  \notag \\
\mathbf{J}_{-\frac{\pi}{4}}(\pi/2) \mathbf{J}_{\theta_1}(\pi) \mathbf{J}_{\theta_1 + \frac{\alpha}{4}}(\pi) \mathbf{J}_{\frac{\pi}{4}}(\pi/2) \mathbf{J}_{\theta_2}(\pi) \mathbf{J}_{\theta_2 + \frac{\beta}{2}}(\pi).
\end{gather}

After fixing the free parameters $\theta_1 = \pi/4$ and $\theta_2 = -\pi/4$ and after some trivial simplifications we obtain the final form of our total propagator:

\begin{equation}
\mathbf{J} = \mathbf{J}_{\frac{\pi}{4}}(\pi/2) \mathbf{J}_{\frac{\pi}{4} + \frac{\alpha}{4}}(\pi) \mathbf{J}_{-\frac{\pi}{4}}(\pi/2) \mathbf{J}_{\frac{\beta}{2} - \frac{\pi}{4}}(\pi) = \mathbf{J}_{0}(\alpha) \mathbf{R}(\beta),  
\label{arbitrary-to-arbitrary polarization}
\end{equation}

\noindent which defines a Jones matrix for the combination of a tunable retarder with a tunable rotator. In general, the proposed optical polarization controller can map any input polarization state to any other polarization state at the output. However, some important restrictions must be mentioned. Detailed examination of Eq. (\ref{arbitrary-to-arbitrary polarization}) shows that not every input state can be mapped to any output polarization state. For example, if the input is circularly polarized and we first apply rotation (as in eq. (\ref{arbitrary-to-arbitrary polarization})) the polarization state does not change after the first rotator and as a result, a very limited number of output states are possible. In order to solve this problem, we must first apply the retarder, i.e. we must apply the inverse Jones matrix of Eq. (\ref{arbitrary-to-arbitrary polarization}), which is the same device, but with elements in reverse order. 
The situation is almost the same if the input is linear. In reverse order, very limited output polarization states are possible, but in direct order, linearly polarized input can be transformed into any desired output. In the more general case of elliptical polarization input, depending on the ellipticity more output polarization states might be available either in direct or in reverse order, but in any case not any output. So, one more degree of freedom should be introduced in order to make the proposed polarization controller universal, i.e. reordering of the wave plates.

\section{Experiment}
\label{experiment}

Schematic of the experimental setup used for experimental proof of the proposed device is shown in Fig.\textbf{~\ref{Setup}} a), while Fig.\textbf{~\ref{Setup}} b), represents a photo of the implementation with short description of the components. The periscope is not an essential part of the controller (that is why it's missing on the schematic) and it was used to adjust the hight of the laser beam.  \color{black} The controller shown in Fig.\textbf{~\ref{Setup}} should be considered as a proof of the concept, rather than as a solution with ultimate performance. We used commercial multi-order wave plates and take no special measures, neither on their alignment nor on the control over the environmental conditions. We estimated the performance of the device by finding the most sensitive parameters influencing its accuracy. These are the rotation angles and the retardation phases of the wave plates. By setting reasonable estimates of their uncertainties, we demonstrate a good agreement between the predicted and measured polarization states. Depending on the application, one may use higher quality or even expensive zero-order wave plates to increase the transmission and the bandwidth, precise rotary mounts to improve the position accuracy, temperature control etc.
\color{black}

The wave plates used in the experiments (Fig.\textbf{~\ref{Setup}}) are multi-order 780 nm plates from Thorlabs and Melles-Griot. They are placed in rotation mounts with a resolution of 2$^{\circ}$.  We defined our coordinate system as shown in Fig. \textbf{~\ref{Setup}} a). The \textit{x} direction is perpendicular to the optical table (we will refer to it as “vertical”), while \textit{y,} that is parallel to the table, will be called “horizontal” and z-direction is along the beam propagation and it is also the optical axis of the setup. Positive angles are defined in the counter-clockwise view in the direction opposite to the propagation. As a source, an extended cavity single-mode diode laser having a central wavelength of 780 nm was used. Its polarization was initially set to vertical (along \textit{x}) and further purified by a high-contrast Glan polarizer. \textit{WP1} is a half-wave plate or a quarter-wave plate (or both), used to create various polarization states that we use as input ($S_{\mathrm{in}}$). \color{black}For the initial alignment all four wave plates A to D were inserted into the optical path at normal incidence with their fast axes orientated along x-direction. All wave plates have transmission of more than 97 \%\ but one of the $\lambda/2$ plates used only has 89 \%, resulting in overall transmission  of the controller of around 82 \%.
\color{black}

The output polarization state was measured with a lab-build polarization analyzer assembled from a $\lambda/4$ wave plate (\textit{WP2}) with a fast axis fixed along x-direction (i.e. vertical) and a thin film analyzing polarizer mounted in a rotation mount with a $5'$ resolution. The intensity passed through the system was then measured with a standard large-aperture photodiode. Using the classical method \cite{Analyser}, with four measurements of the transmitted intensity at different orientations of the polarizer, it is possible to calculate the Stokes vector components.

\begin{figure}[th]
    \centering
    \includegraphics[width=1\linewidth]{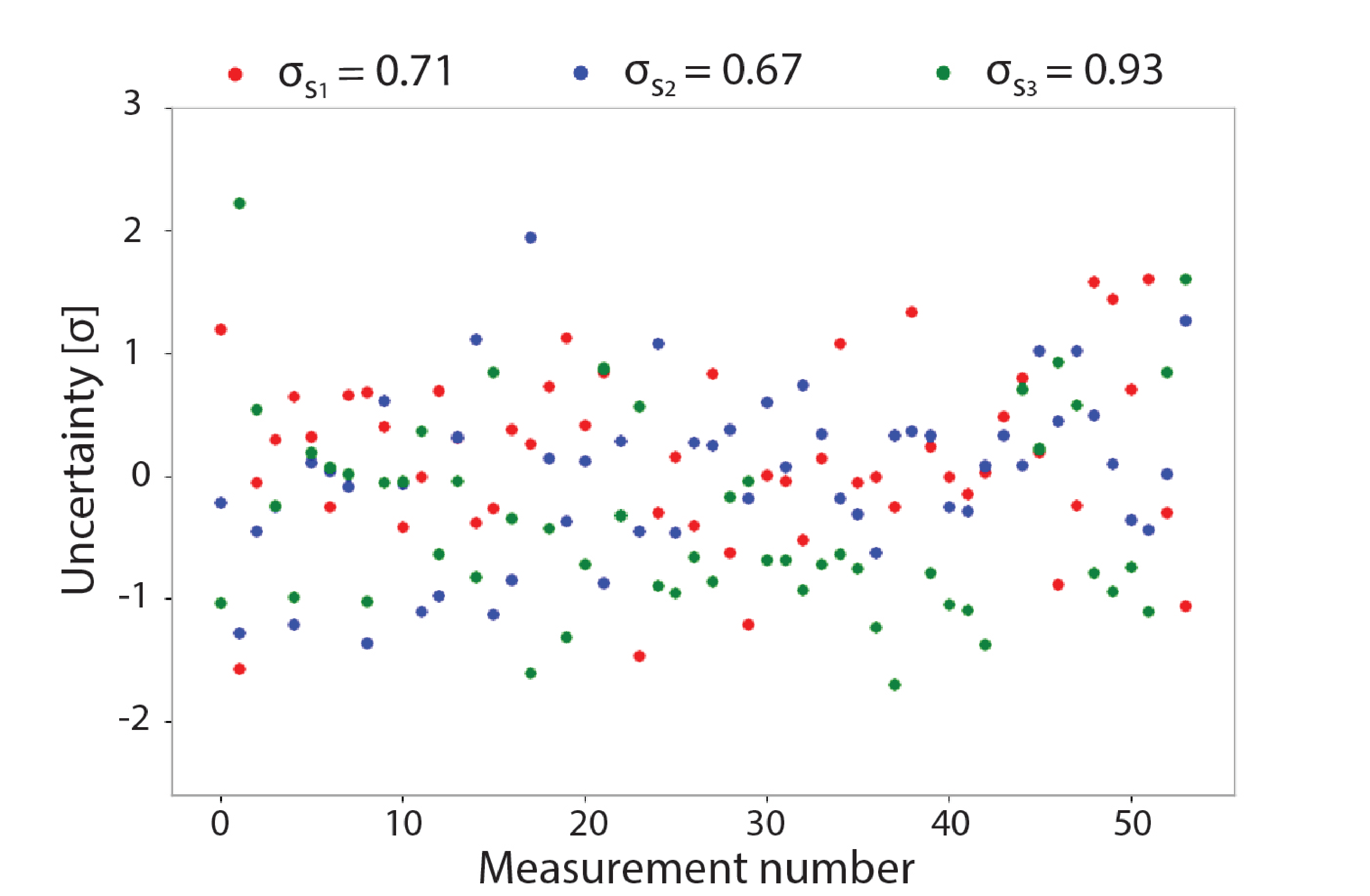}
    \caption{Residuals of the Stokes components for direct order.}
    \label{fig2}
\end{figure}
\begin{figure}[th]
    \centering
    \includegraphics[width=1\linewidth]{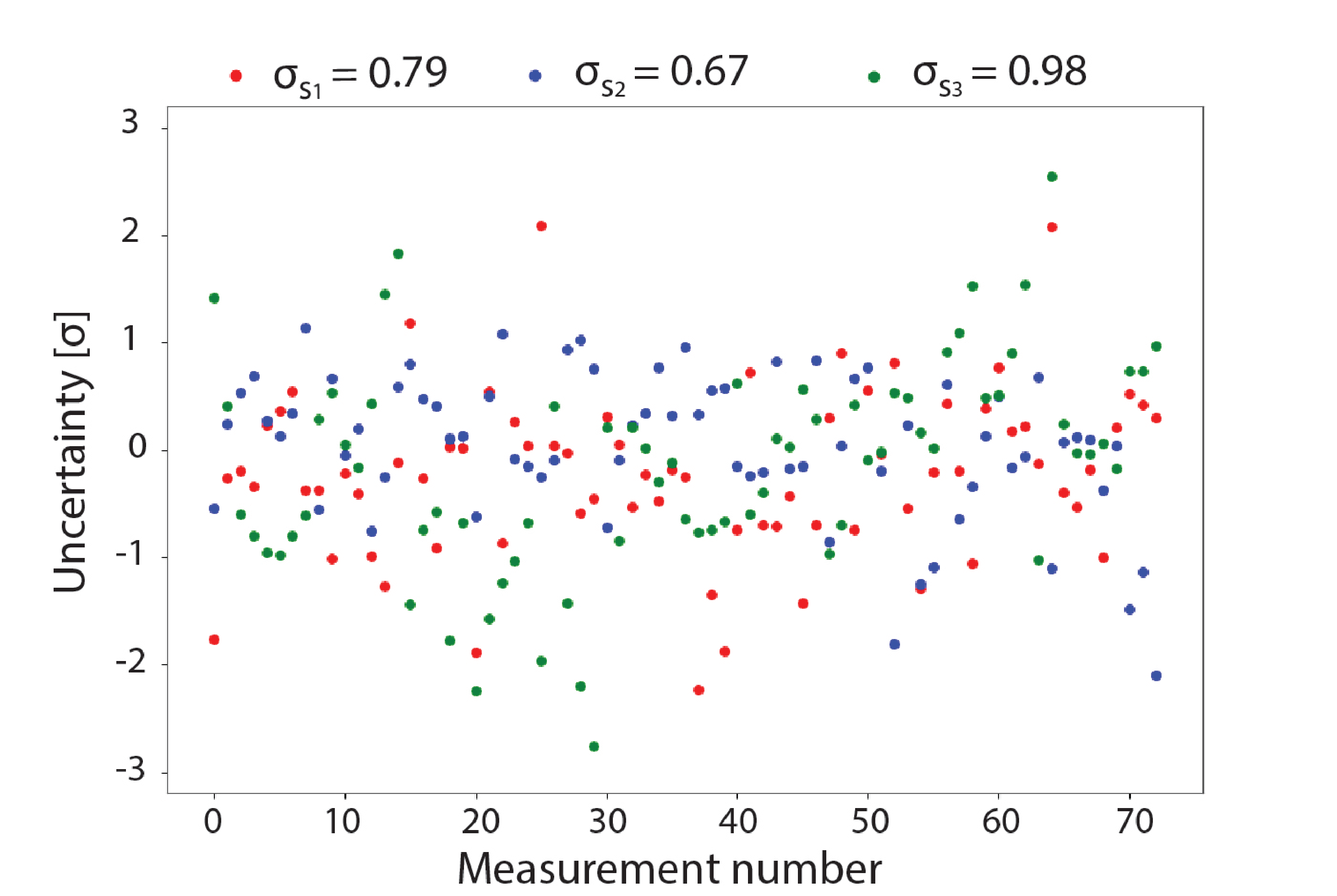}
    \caption{Residuals of the Stokes components for reverse order}
    \label{fig3}
\end{figure}

The operation of the controller was demonstrated in two modes: (i) direct order ABCD, as shown in  Fig.~\ref{Setup} and (ii) reverse order (DCBA). After the initial  alignment, the plates were rotated to $\pm\pi/4$ (see Eq. (\ref{arbitrary-to-arbitrary polarization})), with a precision of $\pm1^{\circ}$, reaching the initial state of the controller for angles $a=\alpha/4=0^{\circ}$ and $b=\beta/2=0^{\circ}$.  We used \textit{WP1} to create different (linear, elliptical, or circular) input polarization states for which the Stokes vector $(S_{\mathrm{in}} \pm \sigma_{\mathrm{in}})$ was experimentally measured. Then the light was passed through the polarization controller with the wave plates A and C rotated at some preset angles $b=\beta/2$ and $a=\alpha/4$ and the output ($S_{\mathrm{out}}^{\mathrm{exp}} \pm \sigma^{\mathrm{exp}}_{\mathrm{out}}$) was measured. Given the angles  $a$, $b$ and the retardation phases $\phi_A...\phi_D$ we calculated the expected output ($S_{\mathrm{out}}^{\mathrm{calc}} \pm \sigma^{\mathrm{calc}}_{\mathrm{out}}$) and compare it with the experimental results. Finally, we plot the normalized residuals:

\begin{equation}
    \frac{S_{\mathrm{out}}^{\mathrm{exp}}-S_{\mathrm{out}}^{\mathrm{calc}}}{\sqrt{(\sigma^{\mathrm{exp}}_{\mathrm{out}})^2+(\sigma^{\mathrm{calc}}_{\mathrm{out}})^2}} \mbox{ .}
    \label{res}
\end{equation}

Since $S_{\mathrm{0}}=1$ (for perfectly polarized light), normalized residuals were calculated only for $S_{\mathrm{1}}$, $S_{\mathrm{2}}$ and $S_{\mathrm{3}}$. A total of more than 130 measurements with various input polarizations were performed. To estimate their uncertainties we took into account the uncertainties in the rotation angles and those of the retardation phases of the wave plates used. For the angles, the uncertainty in the position is associated with the precision of the rotation mounts, which in our case is $1^{\circ}$. The retardation of \textit{WP2} was experimentally measured by illuminating the analyzer with a well-defined linear polarization at various angles and than comparing the expected and the measured Stokes vectors. The real phase of  \textit{WP2} was found to be $87^{\circ}\pm 1^{\circ}$. The phases of the wave plates A to D were assessed in an non-linear fit by minimizing the difference between the predicted and measured output Stokes vectors of all 130 measurements. The final estimations with an uncertainty of $\pm 1^{\circ}$ are $\phi_A=177.6^{\circ}$,  $\phi_B=90.3^{\circ}$, $\phi_C=192.7^{\circ}$ and  $\phi_D=89.1^{\circ}$. \color{black}Within the estimated uncertainties the mechanical stability of the device was sufficient and once adjusted it was capable to operate during a few measurement sessions, meaning at least couple of days, without any additional optical or mechanical alignments.
\color{black}

\begin{table*}
    \centering
    \begin{tabular}{|c|c|c|c|c|c|}
        \hline
         Stokes  & $S_{\mathrm{in}} (\sigma_{\mathrm{in}})$ & Angles & $S^{\mathrm{exp}}_{\mathrm{out}} (\sigma^{\mathrm{exp}}_{\mathrm{out}})$ & $S^{\mathrm{calc}}_{\mathrm{out}} (\sigma^{\mathrm{calc}}_{\mathrm{out}})$ & weighted  \\ 
         components&&&&&residuals (\ref{res})\\ \hline

$S_1$ &   0.995 (0.002) &$\alpha=0^{\circ}$    &  0.929 (0.004) &  0.957 (0.015) &  -1.85  \\
$S_2$ &   0.003 (0.038) &$\beta=0^{\circ}$     & -0.116 (0.030) &  0.004 (0.113) &  -1.03  \\
$S_3$ &   0.009 (0.036) &                      &  0.289 (0.033) &  0.248 (0.046) &   0.72  \\\hline
$S_1$ &   0.995 (0.002) &$\alpha=-211^{\circ}$ &  0.987 (0.004) &  0.989 (0.010) &  -0.14  \\      
$S_2$ &   0.003 (0.038) &$\beta=6^{\circ}$     & -0.097 (0.035) & -0.010 (0.103) &  -0.80  \\      
$S_3$ &   0.009 (0.036) &                      &  0.018 (0.036) & -0.005 (0.067) &   0.30  \\\hline
$S_1$ &   0.995 (0.002) &$\alpha=-179^{\circ}$ & -0.986 (0.002) & -0.989 (0.009) &   0.29  \\      
$S_2$ &   0.003 (0.038) &$\beta=-84^{\circ}$   &  0.048 (0.036) &  0.008 (0.111) &   0.34  \\      
$S_3$ &   0.009 (0.036) &                      & -0.088 (0.037) &  0.001 (0.046) &  -1.52  \\\hline
$S_1$ &   0.995 (0.002) &$\alpha=272^{\circ}$  &  0.071 (0.004) &  0.003 (0.103) &   0.66  \\       
$S_2$ &   0.003 (0.038) &$\beta=-50^{\circ}$   & -0.071 (0.006) & -0.022 (0.116) &  -0.42  \\       
$S_3$ &   0.009 (0.036) &                      &  0.982 (0.013) &  0.984 (0.013) &  -0.09  \\\hline 
$S_1$ &   0.995 (0.002) &$\alpha=-92^{\circ}$  &  0.052 (0.004) &  0.003 (0.110) &   0.44  \\       
$S_2$ &   0.003 (0.038) &$\beta=49^{\circ}$    &  0.076 (0.007) &  0.010 (0.113) &   0.58  \\       
$S_3$ &   0.009 (0.036) &                      & -0.987 (0.004) & -0.983 (0.012) &  -0.28  \\\hline \hline 
$S_1$ &   0.997 (0.001) &$\alpha=45^{\circ}$   &  0.515 (0.009) &  0.464 (0.096) &  0.56      \\       
$S_2$ &  -0.005 (0.021) &$\beta=-33^{\circ}$   &  0.593 (0.017) &  0.690 (0.078) &  -1.17    \\       
$S_3$ &  -0.001 (0.019) &                      & -0.633 (0.015) & -0.528 (0.090) &  -1.11    \\\hline 
$S_1$ &   0.997 (0.001) &$\alpha=120^{\circ}$  & -0.447 (0.006) & -0.469 (0.088) &  0.25     \\       
$S_2$ &  -0.005 (0.021) &$\beta=54^{\circ}$    &  0.410 (0.013) &  0.411 (0.086) &  -0.01    \\       
$S_3$ &  -0.001 (0.019) &                      &  0.778 (0.017) &  0.766 (0.062) &  0.19     \\\hline 
    \end{tabular}
    \caption{Fragment of the experimental results for the controller in direct order and linear input polarization, and several adjustments of the angles $a=\alpha/4$ and $b=\beta/2$. The difference between the measured $S^{\mathrm{exp}}_{\mathrm{out}}$  and the calculated output state $S^{\mathrm{calc}}_{\mathrm{out}}$ is weighted as shown in eq.~\ref{res}.}
    \label{Quant_comp}
\end{table*}

For all measurements the final weighted residuals are shown in Fig.~\ref{fig2} and Fig.~\ref{fig3}. In the case of operation in direct order (Fig.~\ref{fig2}) most of the measurements are in the range $ \pm\sigma$, for $S_{\mathrm{1}}$, $S_{\mathrm{2}}$ and $S_{\mathrm{3}}$  components with standard deviation of the residuals $ \sigma_{\mathrm{S_1}}=0.71$, $ \sigma_{\mathrm{S_2}}=0.67$ and $ \sigma_{\mathrm{S_3}}=0.93$. In the reverse order (Fig.~\ref{fig3}) the uncertainties for some particular measurements exceed $2\sigma$ and reach $3\sigma$. Nevertheless, the standard deviations of the whole set are nearly the same: i.e. $ \sigma_{\mathrm{S_1}}=0.79$, $ \sigma_{\mathrm{S_2}}=0.67$ and $ \sigma_{\mathrm{S_3}}=0.98$.

\begin{figure}[h]
    \centering
    \includegraphics[width=1\linewidth]{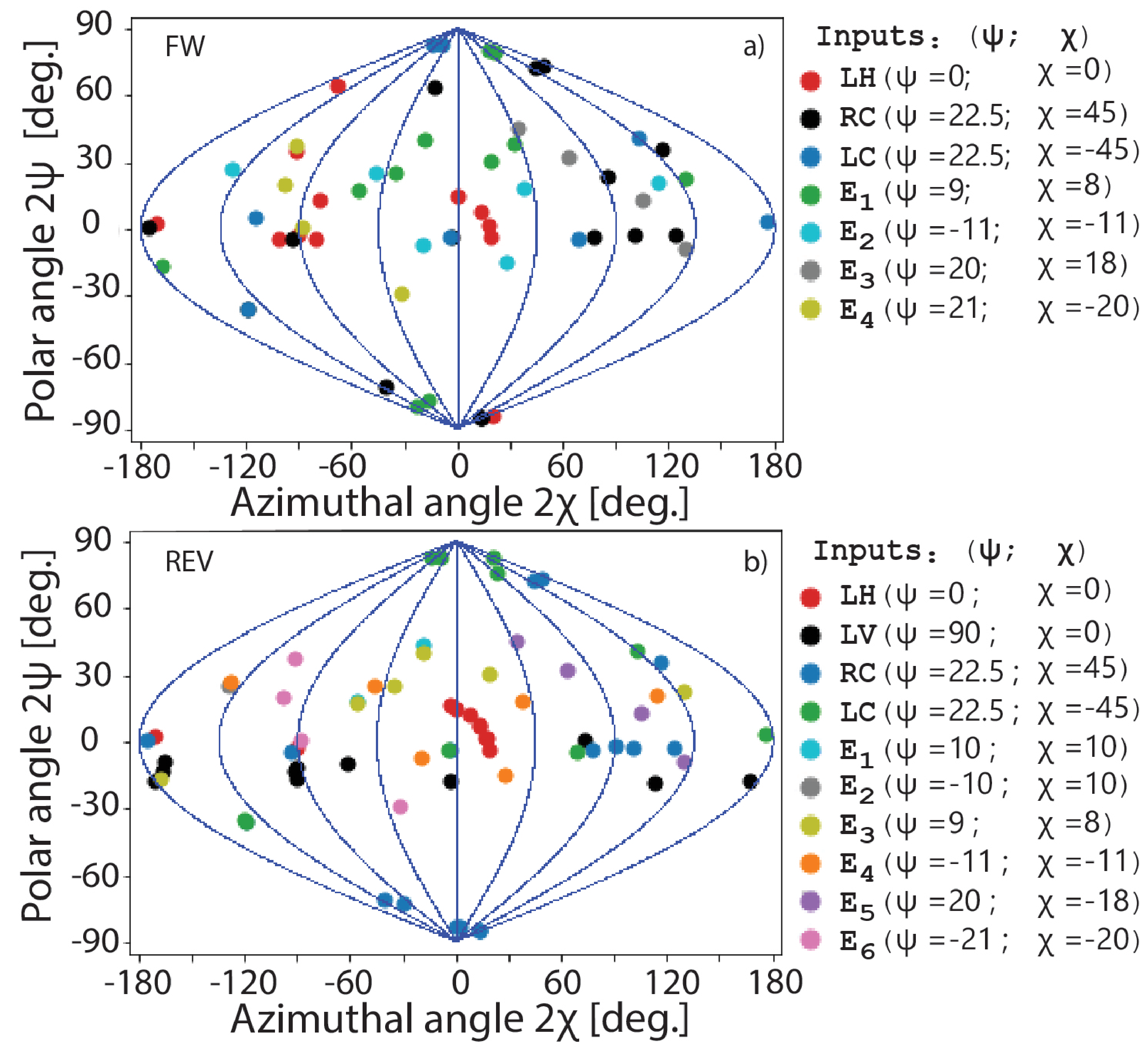}
    \caption{Experimental results for the controller in: a) direct and b) reverse order for different inputs. }
    \label{fig4}
\end{figure}

For perfectly polarized light: 
\begin{equation}  
 {\sqrt{S_1^2+S_2^2+S_3^2}=1} 
\label{pol}
\end{equation}

\noindent so instead of the four Stokes coordinates it is sufficient to use the azimuthal angle $2\psi$ and the polar angle $2\chi$, which define the state vector on the Poincar\'{e} sphere. The same angles, $\psi$, and $\chi$ are used to define the shape of the polarization ellipse \cite{Goldstein}. Each circle in Fig.~\ref{fig4}  represents a measured output state for the controller in direct (Fig.~\ref{fig4} a)) and in reverse order  (Fig.~\ref{fig4} b)), while each color corresponds to different input polarization state respectively; LH - Linear Horizontal, LV - Linear Vertical, LC - Left-hand Circular, RC - Right-hand Circular, E - Elliptical with different ellipticity defined by $\psi$ and $\chi$. 

\color{black}
To produce any desirable output polarization state $S_{\mathrm{out}}$ from given input $S_{\mathrm{in}}$ it is necessary to solve the so-called inverse problem and to find the angles $\alpha$ and $\beta$. This problem is difficult to treat analytically, so we solved it numerically. For the calculations, we used Stokes vectors for the polarization states and M\"uller matrices derived from the Jones matrices. Starting from some initial roughly estimated angles, we calculate numerically the derivatives of the output Stokes components with respect to $\alpha$ and $\beta$. Then we searched iteratively for angles that minimize the difference between $S_{\mathrm{out}}$ and $\mathbf{M}(\alpha,\beta)S_{\mathrm{in}}$ (where $\mathbf{M}(\alpha,\beta)$ is the M\"uller matrix of the controller). If a solution exists, convergence is fast and a few iterations are sufficient. Usually the agreement between the components of the two Stokes vectors is less than a percent. However, the solution of the inverse problem is not unique and depending on the initial angles different solutions are possible, meaning that a single output state can be reached by more than one configuration of the controller. When the initial values for $\alpha$ and $\beta$ are far from the optimal, the agreement may not be satisfactory, and then we can repeat the procedure using the results from the first iteration as input for the next. If solution does not exist (such examples are shown in fig.~\ref{fig6} when transforming elliptical polarization) the algorithm does not converge.     

In order to provide more comprehensive and quantitative metrics for the performance and the accuracy of the proposed controller, a fragment of the experimental results, for linear input polarization is shown in Table~\ref{Quant_comp}. 
The expected outputs ($S^{\mathrm{calc}}_{\mathrm{out}}$) were calculated with the real phases of the plates, for measured input state $S_{\mathrm{in}}$, and also with $\alpha$ and $\beta$ rounded to the nearest integer.
The uncertainties for $S_{\mathrm{in}}$ and $S^{\mathrm{exp}}_{\mathrm{out}}$ result from the accuracy of our polarization analyzer. The uncertainties of $S^{\mathrm{calc}}_{\mathrm{out}}$ are much larger. They account for the uncertain adjustment of the wave plates rotation angles, the uncertainties in the wave plate phases and also for the uncertainty in $S_{\mathrm{in}}$.  

With the first measurement we demonstrate that for $\alpha = \beta = 0$ the output state is significantly different from what one could expect for device with ideal wave plates, i.e. unchanged. This happens because the phases of the used wave plates deviate from the ideal ones, but experimentally measured values ($S^{\mathrm{exp}}_{\mathrm{out}}$) are still in a good agreement with the calculated ($S^{\mathrm{calc}}_{\mathrm{out}}$) ones.

In the second measurement, we solved the inverse problem and we demonstrated that the controller indeed can preserve the input, but at different angles, respectively at $\alpha=-211^\circ$ and $\beta=6^\circ$. As a result the measured output state $S^{\mathrm{exp}}_{\mathrm{out}}$ is in much better agreement with $S_{\mathrm{in}}$.

In the following measurements we aligned the controller to produce horizontal $(1,-1,0,0)$, left circular $(1,0,0,1)$ and right circular $(1,0,0,-1)$ polarizations. 
The last two measurements demonstrate transformations from a vertical linear input to two different elliptical output states, $(1,0.45,0.70,-0.55)$ and $(1,-0.48,0.44,0.76)$ respectively. In all these cases, we achieved a good agreement between the desired ($S^{\mathrm{calc}}_{\mathrm{out}}$)  and  experimentally measured ($S^{\mathrm{exp}}_{\mathrm{out}}$) outputs.

The components of the Stokes vector are very sensitive to the adjustment of the wave plates and their phases, which is visible from the estimated uncertainties $\sigma^{\mathrm{calc}}_{\mathrm{out}}$. We will discuss possible solutions in the last section of the paper.
\color{black}

\section{Simulations}
\label{simulation}

As already mentioned, in some cases the controller cannot map the input to some desirable output state. In order to achieve a more profound understanding of these cases, we carried out a numerical simulation that, for a given input ($S_{\mathrm{in}}$), maps any possible combination of $S_{\mathrm{out}}$ by choosing angles $a\in[0,180]$ and $b\in[0,180]$ in a random way (uniform distribution). 
\begin{figure}
    \centering
    \includegraphics[width=1\linewidth]{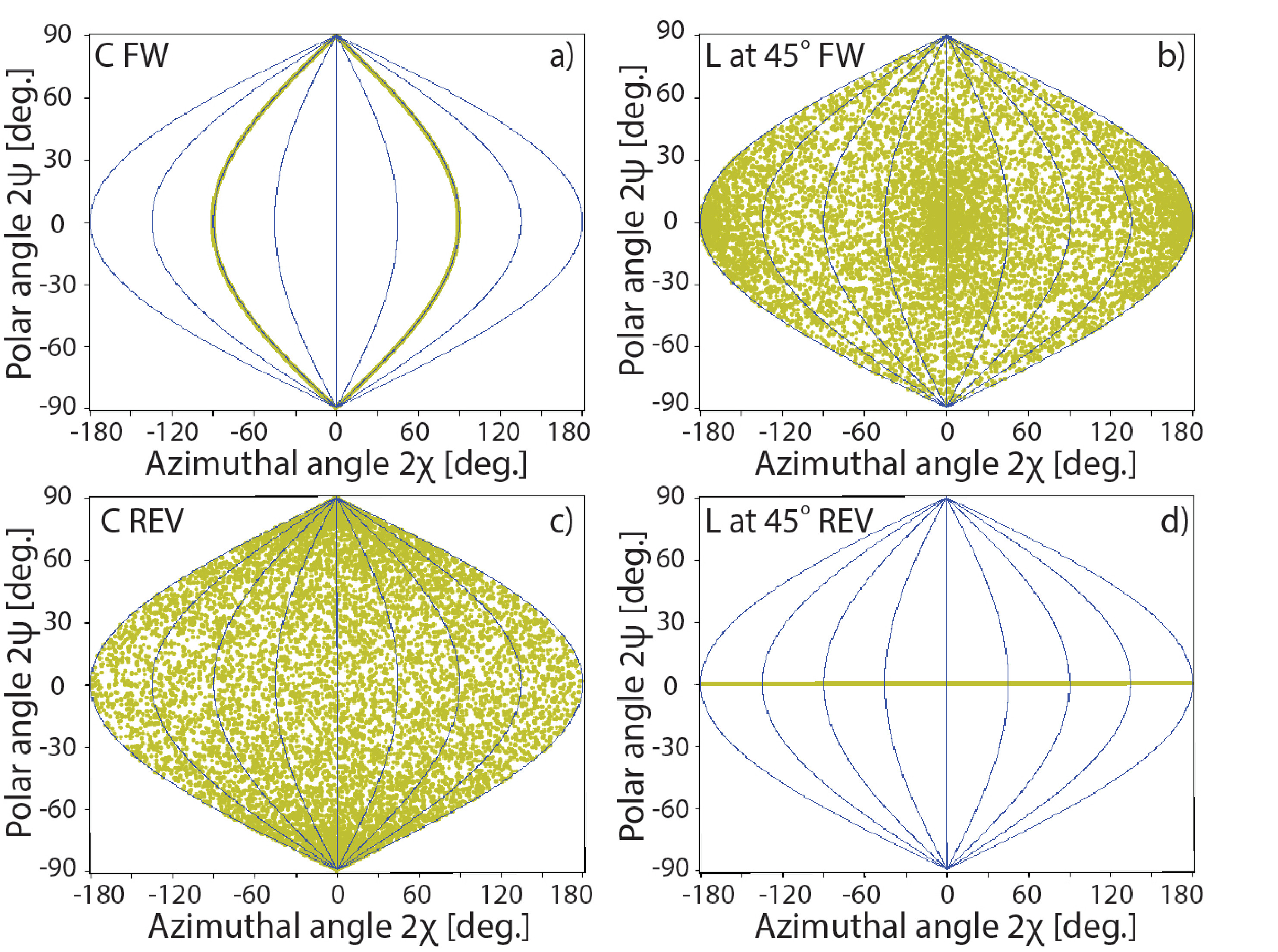}
    \caption{Numerical simulations for: C FW - circular input and direct order, C REV - circular input and  reverse order, L at 45$^{\circ}$ FW - linear input at 45$^{\circ}$ and direct order, L at 45$^{\circ}$ REV - linear input at 45$^{\circ}$ and reverse order}
    \label{fig5}
\end{figure}

In Fig.~\ref{fig5} some results from this modeling are shown as $(2\psi,2\chi)$ output angles for 10000 combinations of $(a,b)$. The input states are linear at $45^{\circ}$ and left circular. Every circle in Fig.~\ref{fig5}, represents a calculated output state while the white space is a combination of Stokes parameters that are missing in our simulation results. It is useful to look at the state's density. The regions where the circles overlap have very high density. These regions are easily reachable and not very sensitive to the precise alignment of the controller. However, the regions where the density is low are more sensitive to alignment but still reachable. Finally, the regions without any output states are not accessible for this input state. In the cases with circular input and the controller in direct order (Fig.~\ref{fig5}, a) a very limited number of output states are available. In the direct order, we first apply rotation which does not change the circular input. After that, we apply retardation, with an axis fixed at $45^{\circ}$, which will result in linear polarization, and what is left for the last two wave plates is to rotate this linear polarization and introduce some retardation resulting in a limited number of elliptical polarization states (the green meridians on  Fig.~\ref{fig5}, a)). On the other hand, with the same circular input, the controller in reverse order can map the input to any output polarization state (Fig.~\ref{fig5}, c). The situation is similar, although exactly the opposite for linear input at $45^{\circ}$. With the controller in direct order (Fig.~\ref{fig5}, b) any possible output will be available while in reverse order (Fig.~\ref{fig5}, d), a very limited number of states around the equator are accessible. For other linear inputs, different from $45^{\circ}$, almost any output state is available in direct as well as in the reverse order.

\begin{figure}[h]
    \centering
    \includegraphics[width=1\linewidth]{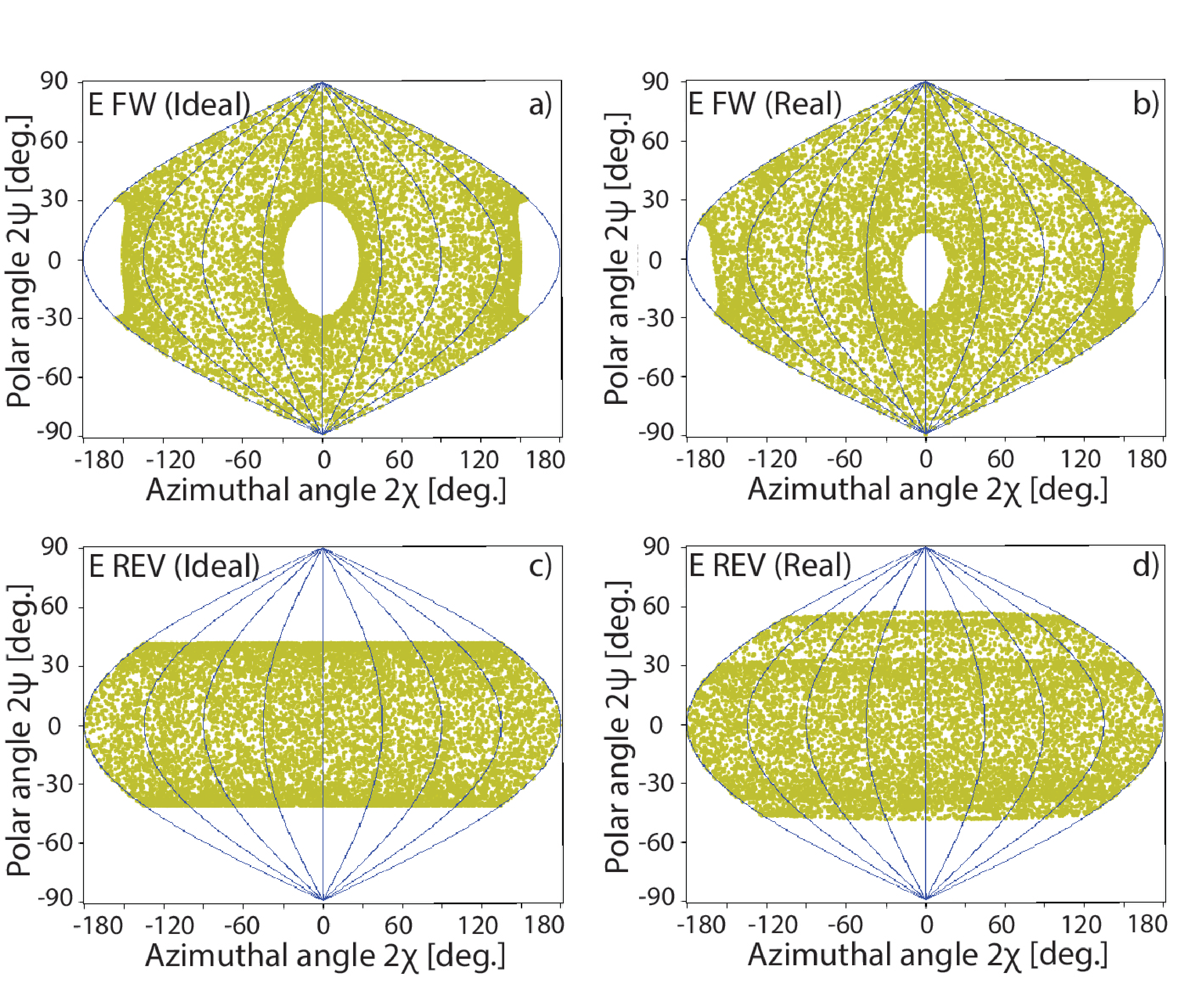}
    \caption{Numerical simulations for elliptical input and controller in direct (FW) and reverse (REV) order. The Ideal and Real refer to the case when ideal (e.g. $\phi=90^{\circ}$ and $\phi=180^{\circ}$) and real (measured) phases are used in the calculations}
    \label{fig6}
\end{figure}
In the case of an elliptical input, which is the most general case, the controller cannot map all possible output states neither in direct nor in reverse order, as shown in Fig.~\ref{fig6}, a) and c). There are clearly distinguishable regions that represent output polarization states that are unreachable to the controller, but there are no output states missing for both directions.  

For comparison, Fig.~\ref{fig6}, b) and d) represent the same numerical calculations, but instead of ideal wave plates we have inserted the estimated values for the phases of our wave plates, and it looks like in our case this is somehow beneficial, because the size of the unreachable regions is reduced. Of course, the area of the regions strongly depends on the phases and in some cases the effect can be negative, making unreachable regions larger.  

\section{Conclusion}

In conclusion, a simple device (Fig.~\ref{Setup}) for polarization control has been introduced theoretically and verified experimentally. Although the theory (\ref{arbitrary-to-arbitrary polarization}) requires exact $\lambda/4$ and $\lambda/2$ retardation phases, we demonstrated that it is possible to use the controller with real wave plates, even multi-order ones, which phases are wavelength dependent. Small deviations from the ideal phases can be compensated by adjusting the angles $a$ and $b$. Of course, that deviation may affect the universality of the proposed controller, and some output states may become inaccessible. Alternatively, in fig.~\ref {fig6} one can see a comparison between the performance of an ideal and a real device, where the range of output states is slightly wider for the real device. In any case, for a given wavelength, the behavior of the controller is fully predictable if the real phases of the wave plates are known. \color{black} Therefore, it can be used in a relatively wide spectral range even with multi-order wave plates if the wavelength dependence of their phases is well known. If the application requires broadband operation, it cannot be achieved with multi-order wave plates, but one may use zero-order, achromat wave plates, or Fresnel rhombs.
\color{black}
We identified the most critical parameters that influence the device's performance. Along with the phases, these are the rotation angles of the wave plates. A possible improvement would be the use of more precise rotary mounts and more precise determination of the retardation phases, including their wavelength dependence. Another solution is to monitor the output state by a fast polarization analyzer and adjust the angles $a$ and $b$, in real time, until the desired output state is reached. One can even automatize this procedure by using motorized rotary mounts coupled through a proportional-integral-differential controller to the polarization analyzer. This may be beneficial when one needs a stable polarization state independent of fluctuations in the input Stokes parameters.

\acknowledgments
This research is partially supported by the Bulgarian national plan for recovery and resilience, contract BG-RRP-2.004-0008-C01 SUMMIT: Sofia University Marking Momentum for Innovation and Technological Transfer, project number 3.1.4.

\end{document}